\documentclass[prl, superscriptaddress, twocolumn]{revtex4}
\usepackage{amssymb}
\usepackage{graphicx}
\usepackage{amsfonts}
\usepackage{amsmath}
\usepackage[english]{babel}

\newcommand{\ket}[1] {\left|#1\right\rangle}
\newcommand{\bra}[1] {\left\langle#1\right|}

\newcommand{\ra}{\rightarrow}

\newcommand{\be}{\begin{equation}}
\newcommand{\ee}{\end{equation}}
\newcommand{\ba}{\begin{array}}
\newcommand{\ea}{\end{array}}

\newcommand{\six}{\sigma^x}
\newcommand{\siy}{\sigma^y}
\newcommand{\siz}{\sigma^z}

\newcommand{\rem}[1]{}

\begin{document} 

\title{Adiabatic quenches through an extended quantum critical region}

\author{Franco Pellegrini}
\affiliation{NEST-CNR-INFM \& Scuola Normale Superiore di Pisa, Piazza dei Cavalieri 7, I-56126 Pisa, Italy}
\affiliation{International School for Advanced Studies (SISSA), Via Beirut 2-4, I-34014 Trieste, Italy}
\author{Simone Montangero}
\affiliation{NEST-CNR-INFM \& Scuola Normale Superiore di Pisa, Piazza dei Cavalieri 7, I-56126 Pisa, Italy}
\author{Giuseppe E. Santoro}
\affiliation{International School for Advanced Studies (SISSA), Via Beirut 2-4, I-34014 Trieste, Italy}
\affiliation{CNR-INFM Democritos National Simulation Center, Via Beirut 2-4, I-34014 Trieste, Italy}
\affiliation{International Centre for Theoretical Physics (ICTP), P.O.Box 586, I-34014 Trieste, Italy}
\author{Rosario Fazio}
\affiliation{International School for Advanced Studies (SISSA), Via Beirut 2-4, I-34014 Trieste, Italy}
\affiliation{NEST-CNR-INFM \& Scuola Normale Superiore di Pisa, Piazza dei Cavalieri 7, I-56126 Pisa, Italy}

\date{\today}

\begin{abstract}
By gradually changing the degree of the anisotropy in a XXZ chain we study the defect 
formation in a quantum system that crosses an extended critical region. 
We discuss two qualitatively different cases of quenches, from the antiferromagnetic to the 
ferromagnetic phase and from the critical to the antiferromegnetic phase. 
By means of time-dependent DMRG simulations, we calculate the residual energy at the end of the quench 
as a characteristic quantity gauging the loss of adiabaticity. 
We find the dynamical scalings of the residual energy for both types of quenches, and compare them
with the predictions of the Kibble-Zurek and Landau-Zener theories.
\end{abstract}

\maketitle

When, by changing some external parameters, a quantum system crosses a phase transition at a 
finite speed it is unable to reach its equilibrium (or ground) state no matter how slow the 
rate of the quench is. The reason is that the anomalous slow dynamics close to the critical 
point prevents the system to follow adiabatically the external drive. As a result, supposing that 
the dynamics takes the system to a symmetry broken phase, a number of defects will appear
in the final state. Dynamical defect formation has important implications in a wide spectrum 
of problems ranging from the study of phase transitions in the early universe~\cite{Kibble_JPA76, Kibble:review} 
and in superfluid systems~\cite{Zurek_NAT85,Zurek_APP93,Zurek:review}, to
quantum annealing~\cite{Finnila_CPL94,Kadowaki_PRE98,Santoro_SCI02,Das_Chakrabarti:book,Santoro_JPA:review} 
and adiabatic quantum computation~\cite{Farhi_SCI01}. 

Early works on the adiabatic crossing of a phase transitions dealt with classical system 
where the external control parameter is the temperature. This problem was more recently 
explored also in the case of a quantum phase transition~\cite{Zurek_PRL05,Polkovnikov_PRB05}.
These works stimulated an intense theoretical activity 
(see~\cite{Damski_PRL05,Dziarmaga_PRL05,
Damski_PRA06,Cherng_PRA06,Schutzhold_PRL06,Cincio_PRA07,Cucchietti_PRA07,Fubini_NJP07,
Polkovnikov_07:preprint,Damski_PRL07,Lamacraft_PRL07,Caneva_PRB07,Sengupta_07:preprint,Damski_07:preprint} and 
references therein). 
The large body of results obtained so far for the adiabatic crossing through a critical point
are in agreement with the Kibble-Zurek (KZ) theory.
At the roots of the KZ mechanism there is the hypothesis that the dynamics 
of a system close to a continuous phase transition can be considered adiabatic or impulsive 
depending on the vicinity to the critical point. The determination of the point in which the 
system stops following the external drive leads to the determination of the scaling of density of 
defects and other observables as a function of the quench rate. 
In the limit of very slow quenches, defect formation can be also understood by means of the Landau-Zener (LZ)
theory applied to the ground and the first excited states.

The passage through a critical point is not the only situation one can envisage in this context. 
Another paradigmatic case is when the evolving many-body system is quantum critical in an 
extended region of the parameter space. This is the topic of the present work. The system we use 
to illustrate this situation is described by the XXZ model~\cite{Takahashi:book}. 
The parameter that will be changed to cross the different phases is the anisotropy coupling. 
The study of adiabatic quenches in the XXZ model allows to test several aspects of the problem 
which could not be addressed previously. The system has an extended critical line 
instead of a critical point (a similar issue was recently considered in~\cite{Sengupta_07:preprint} 
for the Kitaev model). Moreover, the boundaries of the critical region are characterized by different 
exponents, hence one can test if the defect density is controlled by what happens before or after 
the passage through the critical point~\cite{Antunes_PRD06}. Finally, this model allows to study 
defect formation in the presence of dynamical constraints. Because of the conservation of the total 
magnetization, it will be impossible for the system to reach the local ground state no matter how slow 
the quench is even for a finite system. 

\begin{figure}[!t]
\centering
\includegraphics[width=0.45\textwidth]{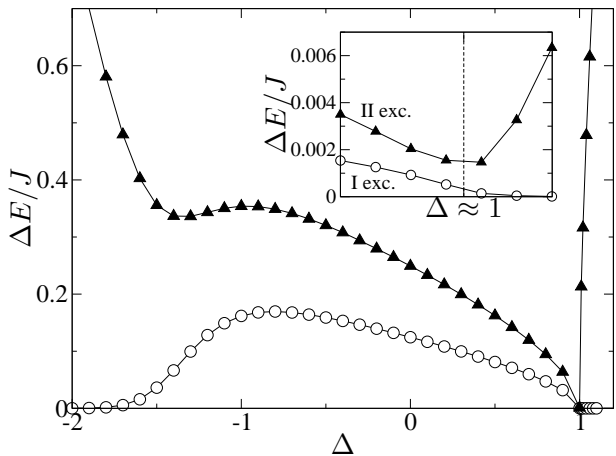}
\caption{\label{Sp100}
         Excitation energy of the two lowest excited states of the XXZ model for $N=100$ 
         spins with open boundary conditions in the subspace $S^z_{\rm tot}=0$. 
         Data obtained from static DMRG simulations ($m=160$ with $3$ target states). 
         Details of the spectrum in the region close to $\Delta=1$ are shown in the inset.}
\end{figure}

The one-dimensional XXZ model is defined by the Hamiltonian
\be 
           H(t) =-J\sum_{i=1}^{N-1}\left[\six_i\six_{i+1}+\siy_i\siy_{i+1} 
                + \Delta(t)\siz_i\siz_{i+1}\right] \;\;.
\ee
describing $N$ spin-$\frac{1}{2}$ interacting via a nearest-neighbour Heisenberg interaction 
anisotropic along the $z$-direction, $\Delta$ being the anisotropy parameter. 
Here $\six$, $\siy$ and $\siz$ are the Pauli matrices. 
This system is invariant under rotations around the $z$ axis, 
so that the total z-component of the spin $S^z_{\rm tot}$ is a conserved quantity. For time-independent 
couplings, the system can be exactly solved, by means of the Bethe ansatz~\cite{Takahashi:book}. 
If $\Delta>1$ the system is ferromagnetic, with 
all spins aligned in the $z$-direction. The low-lying excitations 
have a gap $\Delta E=4J(\Delta-1)$ which closes for $\Delta\ra 1^+$.
For $\Delta<-1$ the system is in the antiferromagnetic N\'eel phase.
The two possible ground states, differing by a traslation 
by one lattice spacing, are degenerate in the thermodynamic limit, i.e., for $N\ra\infty$ 
their splitting is $\propto e^{-c N}$, with $c$ a constant. 
The low-lying excitations are made up by domain walls 
separating regions with the two different N\'eel phases characterizing the ground state: they have 
a total magnetization $S^z_{tot}=0,\pm 1$ and a finite gap which again closes for $\Delta\ra -1^-$. 
The part of the spectrum which is relevant for our pourposes belongs to the $S^z_{\rm tot}=0$ sector.
The excitation energies of the two lowest-lying excited states in this subspace 
for a system of $N=100$ spins and open boundary conditions (OBC) are shown in Fig.~\ref{Sp100}. 
In the whole region $-1\leq\Delta\leq 1$ the spectrum is gapless. 
For finite sizes $N$ and $-1\leq\Delta\leq 1$, the gap  vanishes linearly in $N^{-1}$, 
$\Delta E \approx 2\pi v/N$, for all values of $\Delta$ (with different $\Delta$-dependent velocities $v$), 
except for $\Delta=1$, where the scaling is quadratic, $\Delta E\approx 1/N^2$.
Therefore one expects a final density of defects strongly dependent on whether,
during the quench, the system has crossed this point.

In order to study the time-dependent XXZ model we have to resort to numerical simulations. 
The results we present have been obtained by means of the time-dependent Density Matrix Renormalization 
Group algorithm (t-DMRG) with a second order Trotter expansion of $H$~\cite{Daley_JSTAT04,White_PRL04}
(see~\cite{DeChiararev} for a review). We considered chains up to $N=200$ with open boundary conditions. 
For the slowest quenches we had to restrict $N$ to smaller 
values in order to obtain reliable results. The smallest Trotter time-steps were chosen to be 
$\delta_t = 10^{-4}J$, and the truncated Hilbert space dimension in the DMRG was up to $m=200$. 
We checked that in all the cases presented the results do not depend on the Trotter discretization 
$\delta_t$ and on the DMRG-truncation of the states.

The fully polarized ferromagnetic ground state (i.e., all spins up or down) is a good eigenstate 
for every value of $\Delta$. It is therefore relevant to consider only quenches that start either 
from the antiferromagnetic or from the critical region. 
The anisotropy parameter $\Delta$ is changed in time according to $\Delta(t)= t/\tau_Q$. 
In this work we considered two different situations. 
i) An  evolution from the antiferromagnetic ground state with an initial value of the
anisotropy  $\Delta_{i} \ll -1$ to the ferromagnetic region at a final value $\Delta_{f} \gg 1$; 
ii) an evolution from $\Delta_{i}=0$ in the critical region to the antiferromagnetic region at 
$\Delta_{f} \ll -1$. 

In order to monitor the loss of adiabaticity after the quench, we consider the excess final energy of the 
system relative to the ground state in the given subspace, conveniently rescaled.
More precisely
\be
\label{MyRE} 
               {\tilde E}_{\rm res}(t) =\frac{\bra{\Psi(t)}H(t)\ket{\Psi(t)}-
               \bra{\Psi_{GS}(t)}H(t)\ket{\Psi_{GS}(t)}}
               {\bra{\Psi_0}H(t)\ket{\Psi_0}-\bra{\Psi_{GS}(t)}H(t)\ket{\Psi_{GS}(t)}} \;,
\ee
where $\Psi_0$ is the initial wavefunction, which we take to be the ground state of the initial 
Hamiltonian $H_0=H(t_{\rm in})$, and $\Psi_{GS}(t)$ is the instantaneous ground state of $H(t)$ 
(in the subspace $S^z_{\rm tot}=0$). The denominator normalizes  the excess 
energy to the maximum possible attainable value, corresponding to a wavefunction 
$\ket{\Psi(t)}=\ket{\Psi_0}$ which does not evolve at all (totally impulsive regime). 
When $t\ra t_{\rm fin}$, this quantity approaches a value 
${\tilde E}_{\rm res} = {\tilde E}_{\rm res}(t_{\rm fin})$, which coincides with the final number 
of defects (apart from a constant factor) for $t_{\rm fin}/\tau_Q\gg 1$, as only the $z$-polarization 
of the spins counts, in that limit, in determining the final energy of the system.
${\tilde E}_{\rm res}$ naturally takes into account only the defects formed during the quench, 
and ranges from ${\tilde E}_{\rm res}=1$ for a totally impulsive situation (the wavefunction does not evolve
at all), to ${\tilde E}_{\rm res}=0$ for a fully adiabatic evolution. 

\begin{figure}
\centering
\includegraphics[width=0.47\textwidth]{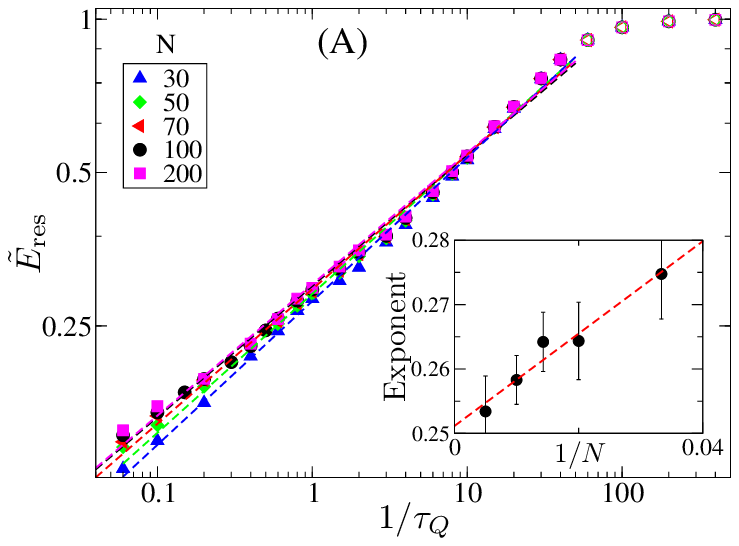}
\vspace{5pt}\\
\includegraphics[width=0.47\textwidth]{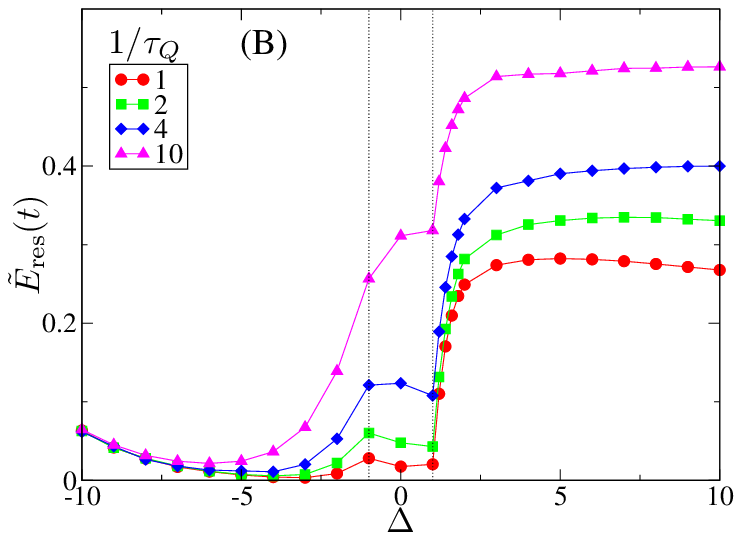}
\caption{\label{FEKZ} (Color online)
        (A) Final excess energy after a quench for XXZ chains of various lengths $N$
        (see legend) as a function of the quench rate $1/\tau_Q$. The dashed lines represent 
        power-law fits of the data for the various $N$'s. 
        Data from t-DMRG simulations ($m$ from $20$ to $30$, $\delta_t$ from $10^{-4}J$ to $10^{-2}J$). 
        Inset: exponents of the power-law fits for various chain lengths $N$ as a function of $1/N$. 
        The dashed line is a linear fit to the exponent, extrapolating to $1/4$ the 
        thermodynamic limit. 
        (B) Evolution of the excess energy of a state during quenches for  
        $N=100$ spins in the region $-10<\Delta<10$ for various rates $1/\tau_Q$. 
        Data from t-DMRG simulations ($m=30$, $\Delta t=10^{-3}J$).}
\end{figure}

{\bf i) Antiferro to Ferro quench} - The system is initially in its ground 
state at $\Delta=-20$ (we tried different initial values, observing no difference), and 
is then quenched at finite rate to a final value of the anisotropy $\Delta=20$. 
The residual energy ${\tilde E}_{\rm res}(t=t_{\rm fin})$ as a function of the quench rate $1/\tau_Q$ 
for various chain lengths is shown in Fig.~\ref{FEKZ}(A).
After a saturation region occurring for very fast quenches, ${\tilde E}_{\rm res}$ obeys a power-law, 
with an exponent which is approximately $0.25$ 
(an extrapolation to the thermodynamic limit would give an exponent $0.251\pm 0.004$, 
see the inset of Fig.~\ref{FEKZ}(A)). The origin of this power-law can be undestood by means  of 
a LZ argument~\cite{Zurek_PRL05}, supposing 
that the loss of adiabaticity is entirely due to the closing of the gap at $\Delta=1$ 
(see Fig.~\ref{Sp100}). The probability of getting excited by this point only is given by
$P_{\rm ex}(\tau_Q,N) = e^{-\gamma \tau_Q (\Delta E_N)^2}$, where $\gamma$ is a constant related to
the slope of the two approaching eigenvalues (ground and excited states), and 
$\Delta E_N=2\pi^2 J/N^2$ is the finite-size smallest gap at the $\Delta=1$ point, 
corresponding to a spin-wave of momentum $k=\pi/N$ (for OBC).
The density of defects for large $\tau_Q$ can be estimated by evaluating
the typical length $\tilde{L}_{\epsilon}(\tau_Q)$ of a defect-free region (in units of the 
lattice spacing $a$),
$\epsilon$ being a small quantity of our choice, denoting the probability of getting excited.
Requiring $P_{\rm ex}(\tau_Q,\tilde{L}_{\epsilon}) = \epsilon$ immediately implies that
$\tilde{L}_{\epsilon}^4=\tau_Q \gamma'/\log{\epsilon^{-1}}$. 
Consequently for the residual energy we get: 
\begin{equation} \label{Eres_estimate:eqn}
\tilde{E}_{\rm res} \sim \frac{1}{\tilde{L}_{\epsilon}(\tau_Q)} \propto \tau_Q^{-\frac{1}{4}} \;,
\end{equation}
in very good agreement with the numerical data. Thus it seems that the scaling of the point $\Delta=1$ 
dominates over the rest of the critical region in a fairly wide region of quench rates.
To understand the importance of this point, and to see how adiabaticity is lost during the quench, 
we looked at the evolution of the residual energy, Eq.~\ref{MyRE}, for various quench rates, see 
Fig.~\ref{FEKZ}(B). What we find is however more complicated than expected.
The wavefunction is not frozen throughout the critical region. 
Nevertheless at $\Delta=1$ there is a clear kink in the departure from 
the adiabaticity which dominates the density of defects (and the residual energy) in the final state.
For the slowest quenches the dependence of the residual energy on the quench rate 
crosses-over from a power-law to an exponential. 
As it has been pointed~\cite{Zurek_PRL05,Damski_PRL05}, this regime is described by the LZ theory.
%
\begin{figure}[!t]\centering
\includegraphics[width=0.48\textwidth]{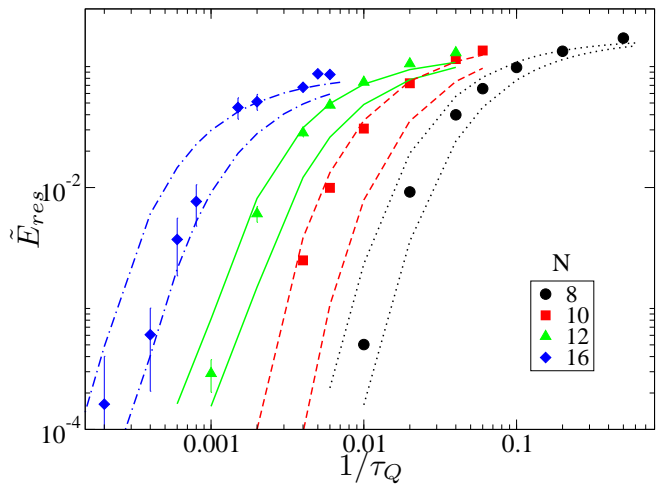}
\caption{\label{FELZ} 
(Color online) Final energy after a quench in XXZ chains of different lengths (see legend) as a function of the quench 
velocity $1/\tau_Q$ (only 'slow' quenches plotted). 
Data from t-DMRG simulations ($m=20$, $\delta_t$ from $10^{-3}J$ to $10^{-2}J$). 
The dashed lines represent LZ prediction starting from the ground state (lower curves) or 
from the first excited state (upper curves, see text), for the various
lengths.}
\end{figure}
%
Here a good estimate of the residual energy can be obtained by assuming that the whole behaviour is 
determined by what happens at $\Delta=1$ where there are two excited states that become degenerate 
at $\Delta \ge 1$. Lower and upper bounds to the residual energy can be simply obtained by considering 
the LZ transition probability from the ground state to the first excited state (lower bound)
and the transition from the first to the second excited state (upper bound). As it can be seen from 
Fig.~\ref{FELZ} the actual rates lie between these two curves. 

{\bf ii) Critical to Antiferro quench} - 
Another interesting situation which can be studied with the XXZ model is the adiabatic quench 
from the gapless phase, for example at $\Delta=0$, to a final point deep inside the N\'eel phase. 
In this case the critical region terminates with a Berezinskii-Kosterlitz-Thouless point at 
$\Delta=-1$, and the applicability of the KZ theory needs to be tested. This situation is 
also relevant for the cases of strongly interacting bosons in one dimension driven from the 
superfluid to the Mott phase~\cite{Schutzhold_PRL06}.

\begin{figure}[!t]
\centering
\includegraphics[width=0.47\textwidth]{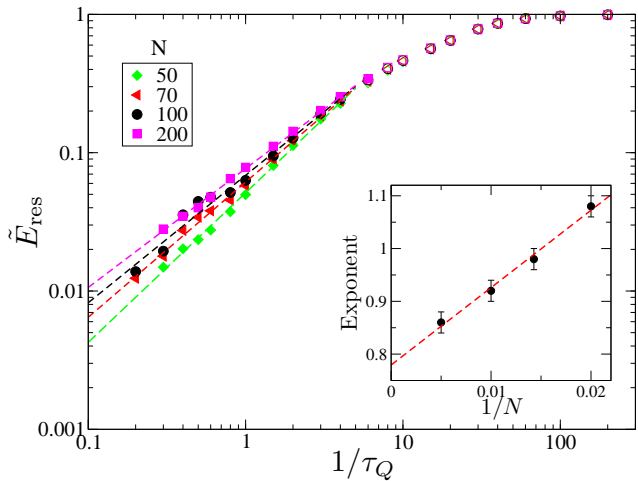}
\caption{\label{Da0a-6} (Color online) Final residual energy after a quench in XXZ chains of different lenght $N$ (see legend) as a 
function of the quench rate $1/\tau_Q$. The dashed lines represent a power-law fit of the data 
(apart from saturation). Data from t-DMRG simulations ($m=30$, $\delta_t=10^{-3}J$). Inset: exponents of the power-law fits for various chain lengths $N$ as a function of $1/N$. The dashed line is a linear fit of the exponents.}
\end{figure}

In our simulations we let $\Delta$ to evolve linearly from $\Delta_i=0$ 
to $\Delta_f=-6$. The residual energy ${\tilde E}_{res}$  as 
a function of the quench rate $1/\tau_Q$ is shown in Fig.\ref{Da0a-6}. The results are apparently 
similar to the previous case. There is a saturation region for very fast 
quenches, which turns into a power-law for slower quenches. The exponent of the power-law decay is 
however different in this case: an extrapolation to the thermodynamic limit of the exponent gives in 
this case a value $0.78\pm 0.02$ 
(see the inset of Fig.~\ref{Da0a-6}).
We believe that this is a true manifestation of the crossing of a critical line. Given the critical 
exponents of the system, a LZ (as well as KZ) treatment of this evolution for 
a single critical point would instead give an exponent $0.5$. 
Differently from the previous case, all the gaps encountered during the evolutions have the same 
scaling and almost the same intensity, as can be seen from Fig.\ref{Sp100}. 
In fact in this region the system undergoes a non-adiabatic evolution through the critical region, 
which is not well described by the models previously proposed. We also note that the arguments recently 
put forward in Ref.~\cite{Sengupta_07:preprint} are not applicable to the present case. 

In conclusion, the XXZ model provides a new paradigm to study adiabatic dynamics 
in many-body critical systems. We showed in this work that depending on the type of quench 
the defect formation is dominated by the presence of a critical point or of a critical line. 
When the quench occurs between the antiferromagnetic and ferromagnetic phase the scaling for the 
defect density can be obtained by focusing on the loss of adiabaticity at $\Delta=1$. In the 
other case we considered, when the quench starts from the critical region and ends in the 
antiferromagnetic region, the exponent obtained signals the crossing of a critical region.  

This research was partially supported by MIUR-PRIN, EC-Eurosqip and CRM {\it ``Ennio De Giorgi''} 
of Scuola Normale Superiore. This work has been developed by 
using the DMRG code released within the `Powder with Power' project (www.qti.sns.it).

%

%
\end{document}